\documentclass[journal,twoside,web]{ieeecolor}
\usepackage{generic}
\usepackage{cite}
\usepackage{amsmath,amssymb,amsfonts}
\usepackage{algorithmic}
\usepackage{graphicx}
\usepackage{textcomp}
\def\BibTeX{{\rm B\kern-.05em{\sc i\kern-.025em b}\kern-.08em
    T\kern-.1667em\lower.7ex\hbox{E}\kern-.125emX}}

\def\w{{\mathbf w}}

\def\1{{\mathbf 1}}

\def\U{{\mathbf U}}

\def\V{{\mathbf V}}

\DeclareMathOperator*{\argmin}{arg\,min}

\begin{document}
\title{High-Dimensional MR Reconstruction Integrating Subspace and Adaptive Generative Models}

\author{Ruiyang Zhao, \IEEEmembership{Student Member, IEEE}, Xi Peng, \IEEEmembership{Member, IEEE}, Varun A. Kelkar, Mark A. Anastasio, \IEEEmembership{Senior Member, IEEE} and Fan Lam, \IEEEmembership{Senior Member, IEEE}
\thanks{
This work was supported in part by the following grants: NSF, NSF-CBET, Grant/Award Number: 1944249. NIH, NIBIB, Grant/Award Number: 1R21EB029076A. }
\thanks{Ruiyang Zhao and Varun A. Kelkar are with the Department of Electrical and Computer Engineering and Beckman Institute for Advanced Science and Technology, University of Illinois Urbana-Champaign, IL 61801, USA. \\
\indent Xi Peng is with the Department of Radiology, Mayo Clinic, Rochester, MN 55902, USA. \\
\indent Mark A. Anastasio and Fan Lam are with the Department of Bioengineering and Beckman Institute for Advanced Science and Technology, University of Illinois Urbana-Champaign, IL 61801 USA (e-mail: fanlam1@illinois.edu).}}
\maketitle

\begin{abstract}
\textit We present a novel method that integrates subspace modeling with an adaptive generative image prior for high-dimensional MR image reconstruction. The subspace model imposes an explicit low-dimensional representation of the high-dimensional images, while the generative image prior serves as a spatial constraint on the ``contrast-weighted" images or the spatial coefficients of the subspace model. A formulation was introduced to synergize these two components with complimentary regularization such as joint sparsity. A special pretraining plus subject-specific network adaptation strategy was proposed to construct an accurate generative-model-based representation for images with varying contrasts, validated by experimental data. An iterative algorithm was introduced to jointly update the subspace coefficients and the multi-resolution latent space of the generative image model that leveraged a recently developed intermediate layer optimization technique for network inversion. We evaluated the utility of the proposed method in two high-dimensional imaging applications: accelerated MR parameter mapping and high-resolution MRSI. Improved performance over state-of-the-art subspace-based methods was demonstrated in both cases. Our work demonstrated the potential of integrating data-driven and adaptive generative models with low-dimensional representation for high-dimensional imaging problems.
\end{abstract}


\section{Introduction}

High-dimensional imaging problems emerge in various MR imaging scenarios, including quantitative MR parameter mapping (qMR)\cite{ma2013magnetic}, dynamic imaging \cite{earls2002cardiac}, and MR spectroscopic imaging (MRSI) \cite{posse2013mr,bogner2021accelerated}, which introduce additional dimension(s) to encode and decode tissue properties and provide quantitative biomarkers for pathological analysis. While the introduction of additional encoding dimensions enriches the available information, it also significantly lengthens the acquisition time, a major challenge for widespread clinical applications. In recent years, reducing the number of encodings and reconstructing images from sparsely sampled data using different model constraints have been recognized as a common approach to accelerate high-dimensional imaging acquisitions. 

Subspace imaging/low-rank tensor modeling is one of the most investigated approaches to enable accelerated high-dimensional MRI. While many variants have been presented, the essence is to exploit the redundancy in the imaging data due to the partial separability property that exists across various high-dimensional imaging modalities \cite{liang2007spatiotemporal,haldar2010spatiotemporal,zhao2012image}. Partial separability can be induced by the spatial-temporal correlations in dynamic MRI \cite{gupta2001dynamic,fu2015high,Lingala_TMI2011_ktSLR}, spatial-parametric correlations in qMR \cite{petzschner2011fast,zhao2015accelerated,zhang2015accelerating}, and spatial-spectral correlations in MRSI \cite{lam2014subspace}, which lead to accurate low-rank representations of the multidimensional imaging data arrays. As a result, high-dimensional image reconstruction can be transformed into the recovery of a low-rank matrix/tensor which requires a significantly fewer number of measurements.

Low-rank modeling can be realized either implicitly through low-rankness encouraging regularization of high-dimensional matrices or tensors (e.g., nuclear norm as in \cite{Lingala_TMI2011_ktSLR,Trzasko_SPIE2013_LLR,qu2015accelerated, Zhang_MRM2015_LLR}), or explicitly through a matrix factorization form that directly reduces the number of unknowns during image reconstruction (e.g., \cite{Haldar_SPL2009_LRPF,zhao2012image,lam2014subspace}). Explicit low-rank modeling represents the signal evolution at each voxel as a linear combination of a small number of ``temporal basis'' with voxel-dependent ``spatial coefficients'' and enables effective incorporation of other prior knowledge to facilitate image recovery. In particular, special hybrid acquisition and physics-driven subspace learning strategies tailored for different applications can be designed to predetermine the basis \cite{christodoulou2014improved,christodoulou2018magnetic, lam2016high}. With the predetermined basis, complementary spatial constraints can be introduced to improve the estimation of the spatial coefficients. For example, analytical sparsifying transforms have been frequently used, e.g., \cite{zhao2015accelerated,peng2016accelerated,christodoulou2013high}. Kernel-based methods have been proposed to leverage image features derived from complementary anatomical imaging modalities to re-represent the spatial coefficients, going beyond the traditional voxel basis paradigm \cite{wang2014pet,li2018constrained}. However, these hand-crafted ``priors'' did not fully exploit the prior information about the unknown images and required carefully feature/kernel selection which can be subjective and sometimes biased. 

On a parallel line of pursuit, deep generative image models have provided a new and flexible way to utilize data-driven prior for image representation \cite{creswell2018generative,tschannen2018recent}, complementary to subspace modeling. State-of-the-art generative models, such as generative adversarial networks (GANs) \cite{aggarwal2021generative}, allow for automatic learning of feature-based image representations, to generate high-quality, high-resolution images and to constrain image reconstruction. For example, the unknown image can be modeled by a pretrained GAN (using images with similar contrasts and resolutions), and the reconstruction problem can be formulated as estimating a set of latent variables input to the GAN instead of the voxel values \cite{bora2017compressed,hussein2020image}. Major challenges of this approach are (1) representation accuracy, i.e., even with fully sampled images as the data, the pretrained GAN may not be able to accurately represent the images; (2) optimization issue, i.e., even if the model is highly expressive, GAN inversion is solving a deep network inversion problem which is highly non-convex and sensitive to initialization. 

Different models combined with various optimization strategies have been proposed to address the above mentioned issues. Asim et al. proposed to use the invertible neural networks (INNs) to mitigate the representation error \cite{asim2020invertible}. As INNs use latent vectors with the same dimension as the unknown image, additional regularization can be introduced to improve reconstruction \cite{kelkar2021compressible}. Image-adaptive GAN (IAGAN) based strategies that update both the lower-dimensional latent vector and weights \cite{hussein2020image,bhadra2020medical} can reduce data-specific representation error (sometimes referred to as ``out of distribution'' error) but may lead to overfiting (e.g., to noise and other artifacts). While untrained networks can also be used as an image model (e.g., deep image prior \cite{ulyanov2018deep,yoo2021time}) for updating only weights during reconstruction, such strategies do not fully utilize data-driven prior available in existing datasets, and the effectiveness highly depends on the network structure.   StyleGAN offers better representation power with flexible latent space adjustments to account for data distribution variations, which demonstrates great potential in accurately representing images and effectively constraining the reconstruction\cite{karras2020analyzing,kelkar2021prior}.

In this work, we proposed a new method that integrates subspace model and an adaptive generative image prior for reconstructing high-dimensional MR images. Direct application of generative models as a spatial constraint for subspace imaging can be challenging. Besides the representation accuracy and network inversion issues, training generative models typically requires large-scale, high-resolution, fully sampled datasets, which can be rather difficult to obtain for applications like qMR, dynamic MRI and MRSI etc. To address these issues, we introduced a pretraining plus subject-specific adaptation strategy for learning an accurate StyleGAN-based image representation. This adaptive prior accounted for contrasts and geometry variations between existing database and application-specific data effectively. Our method does not require end-to-end supervised training using high-dimensional MR data since the GAN can be pretrained on publicly available data, and then adapted to a subject-specific reference image to account for geometry variations. The geometry-adapted GAN representation was incorporated as a spatial constraint into a subspace-based reconstruction formalism. An alternating minimization algorithm was introduced to solve the optimization of jointly updating the spatial coefficients and GAN latent space. We demonstrated the effectiveness of our method using qMR and MRSI data as application examples.

The remaining of the paper is organized as follows: Section II provides some background information on subspace imaging and generative model based image reconstruction. Section III describes the proposed problem formulation and algorithm in details. Section IV presents results evaluating the utility of the proposed method in two application examples, accelerated parameter mapping and MRSI. Section V and VI provide some technical discussion and conclude the paper.

\section{Background}
The acquisition process (after proper discretization) for many high-dimensional MRI problems can be defined as:
\begin{equation}
\mathbf{y}=\mathbf{A} \boldsymbol{\rho}+\mathbf{n},  \label{forwarmodel1}
\end{equation}
where $\boldsymbol{\rho}\in\mathbb{C}^{N \times T}$ corresponds to the discretized representation of the underlying high-dimensional image function with $N$ being the number of voxels and $T$ the number of ``time'' points to be determined, $\mathbf{y} \in \mathbb{C}^{M \times T'}$ denotes the measured data ($T'$ may not necessarily be the same as $T$), $\mathbf{A}$ and $\mathbf{n}$ represent the encoding operator and measurement noise, respectively. The goal is to recover $\boldsymbol{\rho}$ from measurements $\mathbf{y}$ which can be incomplete and/or noisy.

\subsection{Subspace Reconstruction}
While direct reconstruction of the high-dimensional
$\boldsymbol{\rho}$ can be rather challenging and sensitive to noise perturbation, low-dimensional subspace models offer a way to significantly reduce the dimensionality of the problem by exploiting ``correlations" among different dimensions, enabling image recovery from highly sparsely sampled and/or noisy data. More specifically, if $\boldsymbol{\rho}$ is a Casorati matrix \cite{liang2007spatiotemporal} with spatial dimensions cascaded along the first dimension and a temporal/parametric dimension along the second, the subspace model can be realized as a matrix factorization form: $\boldsymbol{\rho} \approx \mathbf{U}\mathbf{V}$, where $\mathbf{U}\in\mathbb{C}^{N \times R}$ and $\mathbf{V}\in\mathbb{C}^{R \times T}$ are low-rank matrices. $\mathbf{U}$ is often referred to as the spatial coefficient matrix and $\mathbf{V}$ as the temporal basis matrix. The model order/rank $R$, is typically much smaller than the ambient image dimensions ($N$ and $T$), which means that the number of unknowns/degree-of-freedom (DOF) is significantly reduced. 

With the low-rank model, the image reconstruction problem can be formulated as
\begin{equation}
\hat{\mathbf{U}},\hat{\mathbf{V}}=\arg \min _{\mathbf{U},\mathbf{V}} \left\|\mathbf{y}-\mathbf{A}(\mathbf{U}\mathbf{V})\right\|_2^2 + \lambda R(\mathbf{U},\mathbf{V}),
\label{eq:direct_lowrank_recon}
\end{equation}
\noindent where $R(\cdot)$ is a spatial/spatiotemporal regularization that incorporates prior information such as sparsity \cite{zhao2012image,Lingala_TMI2011_ktSLR}, piecewise smoothness \cite{yao2018efficient,shi2015lrtv} or structured low rankness \cite{jacob2020structured} etc. This regularization, often including spatial constraints, plays a critical role for the success of subspace reconstruction, as it provides a complementary prior and effectively addresses the instability/ill-conditionness issues related to subspace fitting. Furthermore, for many imaging applications, the temporal basis $\mathbf{V}$ can be pre-determined using a problem-specific hybrid data acquisition strategy \cite{zhao2012image,fu2015high,christodoulou2018magnetic,christodoulou2014improved,lam2016high,zhao2015accelerated}. As a result, the problem in Eq.~\eqref{eq:direct_lowrank_recon} can be simplified to only estimating $\mathbf{U}$, circumventing the nonconvex problem of jointly solving for $\mathbf{U}$ and $\mathbf{V}$. 

\subsection{Generative Model Constrained Image Reconstruction}
Recent advances in deep generative models provide a new way to incorporate domain/data-specific prior into image reconstruction problems. With their state-of-the-art high-quality image generation performance, image representations using convolutional networks, GAN and extensions have been used in inverse problems in imaging. Specifically, a common approach is to represent the unknown image of interest $\mathbf{x}$ as $\mathbf{x} = G_{\boldsymbol{\theta}}(\mathbf{w})$, where $G$ is a generative network (can be pretrained using available domain-specific images or untrained), $\boldsymbol{\theta}$ contains the network parameters, and $\mathbf{w}$ is a set of latent variables (often sampled from i.i.d distributions) that typically has a lower dimensionality than $\mathbf{x}$. With this explicit generative model-based image function, the reconstruction problem can be formulated as
\begin{equation}
\mathbf{\hat{w}},\boldsymbol{\hat{\theta}}=\arg\min _{\mathbf{w},\boldsymbol{\theta}} \left\|\mathbf{y}-\mathbf{A}(G_{\boldsymbol{\theta}}(\mathbf{w}))\right\|_2^2, 
\label{eq:GAN_inverse_bg}
\end{equation}
\noindent which seeks to recover the image by determining the latent $\mathbf{w}$ and/or $\boldsymbol{\theta}$ instead of solving voxel values in the conventional paradigm. This approach assumes that the unknown image lies in the range space of $G_{\boldsymbol{\theta}}$, which may yield substantial modeling errors, depending on the network structure and capacity. This representation accuracy issue has been observed in the early works of compressed sensing using generative model (CSGM) \cite{bora2017compressed}, where a pretrained Deep Convolutional GAN (DCGAN) was used and the reconstruction process only updated $\mathbf{w}$. IAGAN methods \cite{hussein2020image,bhadra2020medical} addressed this issue by simultaneously updating $\mathbf{w}$ and the network parameters $\boldsymbol{\theta}$ to fit the data in Eq.~\eqref{eq:GAN_inverse_bg}. Early stopping is required to avoid overfitting (e.g., to the noisy measurements), particularly when the network has a high capacity. But determining the stopping criterion is heuristic, can be problem dependent and rather challenging.  

The development of advanced generative models, such as StyleGAN2, has demonstrated superior image representation power than early generations of GANs, particularly for high-resolution images \cite{karras2020analyzing}. This makes it a powerful tool in image generation and representation. Many works have been proposed to utilize this representation power for image processing and reconstruction. For example, Kelkar et al. leveraged the multi-resolution latents of StyleGAN2 by constraining the reconstructed images to have similar features with a reference image\cite{kelkar2021prior}. In the meantime, the high-dimensional, nonconvex network inversion problem associated with image reconstruction using GAN-based representation remains challenging and computationally expensive. There are also application specific challenges for high-dimensional MR imaging problems that are in need of formulation and algorithmic innovation.

\section{Proposed Method}
\subsection{Problem Formulation}

We propose to integrate a generative image model as a spatial constraint and the subspace model for high-dimensional image reconstruction via the following formulation:
\begin{equation}
\begin{split}
\hat{\mathbf{U}}, \{\hat{\mathbf{w}}_t\}= & \arg \min _{\mathbf{U}, \{\mathbf{w}_t\}} \left\|\mathbf{y}-\mathbf{A}(\mathbf{U} \hat{\mathbf{V}})\right\|_2^2 \\ & +\sum_{t=1}^{N_t} \lambda_{1,t} \left\|(\mathbf{U} \hat{\mathbf{V}})_t-\boldsymbol{G}_{\hat{\boldsymbol{\theta}}}(\mathbf{w}_t)\right\|_F^2 +\lambda_2 R(\mathbf{U} \hat{\mathbf{V}}).
\label{General_recon}
\end{split}
\end{equation}
The first term in Eq.~\eqref{General_recon} is the well-established subspace/low-rank model with the final reconstruction formed as $\boldsymbol{\rho}=\hat{\U}\hat{\V}$ and $\hat{\V}$ being the predetermined ``temporal ''basis. The second term introduces a GAN-based image representation as a spatial regularization for images at different time points ($t=1,2,..., N_t$). The key assumption is that using a properly trained and adapted StyleGAN with fixed parameters in $\hat{\boldsymbol{\theta}}$, the contrast variations in the image sequence can be accurately accounted for by updating only a set of low-dimensional latent space variables $\{\mathbf{w}_t\}$ (referred to as the latents below; Note that $\mathbf{w}_t$ denotes the entire multi-resolution latent variable for the $t$th image not a particular segment of the latents as described in \cite{kelkar2021prior}). This effectively serves as a data-adaptive reference image to constrain the spatial variations. An updating strategy to explore the multi-resolution latent structure and impose similarity for $\{\mathbf{w}_t\}$ at neighboring time points will be specified in the Algorithm section. The final term $R(.)$ is a hand-crafted regularization term, e.g., the commonly used sparsity constraint \cite{zhao2012image,zhao2015accelerated}. The subspace, sparsity and GAN prior play complementary roles to enhance the reconstruction performance.

The key challenges to integrate a GAN-based image prior are: 1) The representation accuracy of the GAN needs to be validated in application-specific context; 2) For many high-dimensional imaging applications, there may be a lack of high-resolution, high-SNR images for training GAN; and 3) Even with a sufficiently accurate representation, solving the non-convex GAN inversion problem can be quite challenging. To address these issues, we proposed a pretraining plus subject-specific adaptation strategy to construct a StyleGAN prior $\boldsymbol{G}_{\hat{\boldsymbol{\theta}}}(\cdot)$ adaptive to different imaging contexts, leveraging reference images readily available in a specific experiment. We also used an intermediate layer optimization (ILO) algorithm to address the GAN inversion challenge \cite{daras2021intermediate}. More details are provided in the subsequent sections.

\begin{figure}[!t]
\centerline{\includegraphics[width=\columnwidth]{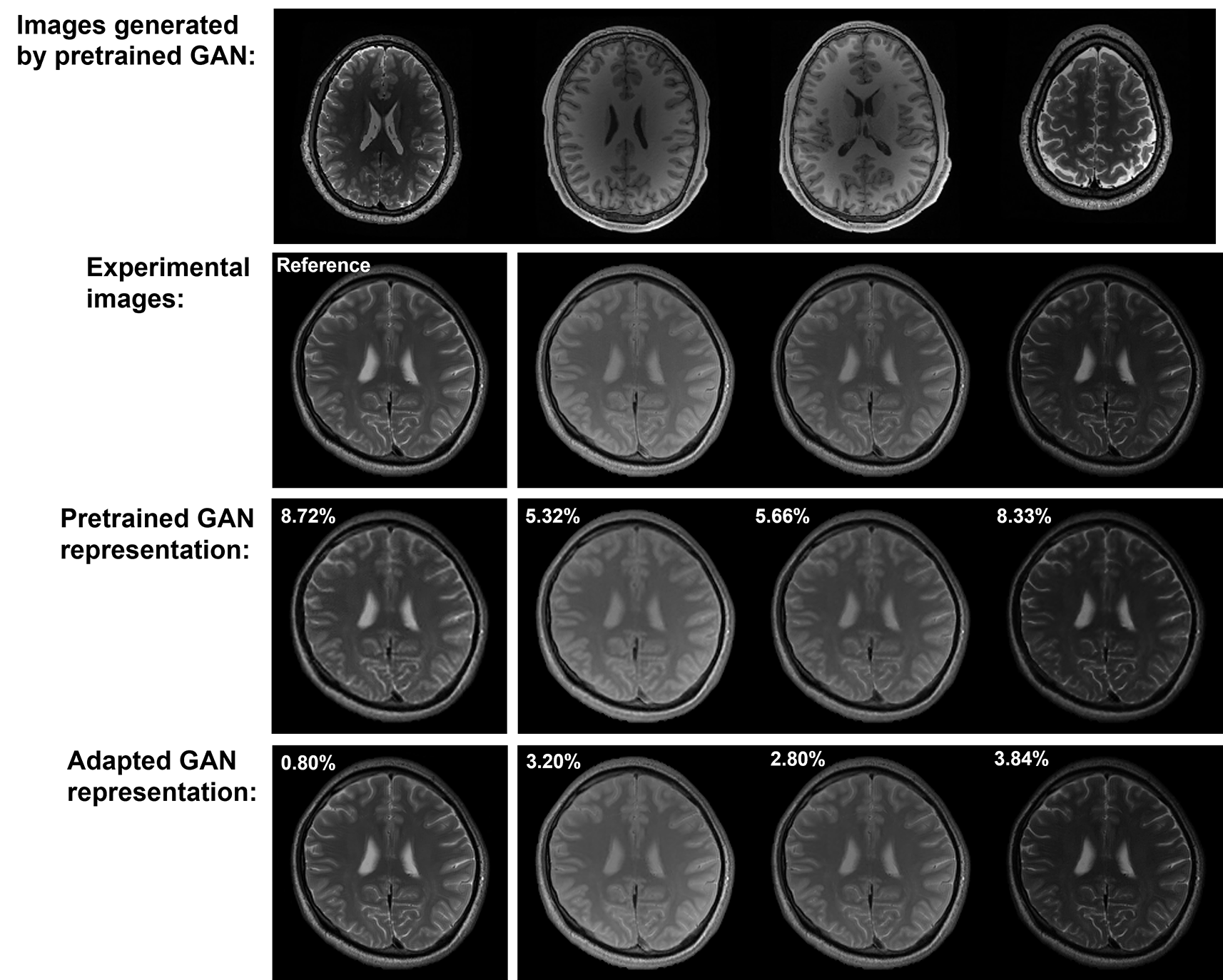}}
\caption{ Validation of the proposed adaptive StyleGAN representation: (Row 1) The pretrained StyleGAN2 can generate high-quality
brain images with different geometries and contrasts; (Row 2) Representative images with different contrast weightings from an independent experiment (not from the database). The first image served as a reference for the GAN adaptation (see texts); (Row 3) Images from the range space of the pretrained StyleGAN2 that were closest to the images in Row 2 in terms of $l_2$ error, from updating the latents of the StyleGAN2. Noticeable representation errors can be observed; (Row 4) The adapted GAN achieved significantly better representation accuracy. Specifically, updating both latents $\w$ and network parameters $\theta$ (left) yielded an image almost the same as the reference image. With the adapted subject-specific network, updating latents only accurately accounted for contrast variations (right three figures in Row 4). These results support using the adaptive StyleGAN2 as an accurate and flexible image representation in a constrained reconstruction.}
\label{Figure_adapt}
\end{figure}

\subsection{Adaptive Generative Image Model}
While it is challenging to directly train a generative model that can accurately capture the geometry and contrast variations in every high-dimensional imaging problems (in many cases such large datasets are not available for training generative networks), a transfer learning plus application-specific adaptation strategy can be developed. Specifically, for a specific organ of interest, we can first train a StyleGAN2 model on large, publicly available data, e.g., the Human Connectomme Project (HCP) database for brain imaging applications \cite{van2013wu}. We chose StyleGAN2 due to its state-of-the-art performance in generating high-quality, high-resolution images, compared to older generations of DCGANs, and its superior image representation capability due to higher dimensional multi-resolution latents.

With the pretrained model, parameterized by $\boldsymbol{\theta}_p$, we proposed a subject-specific adaptation strategy to construct the generative image representation for a specific reconstruction task. The key assumptions are 1) there exists a high-quality reference image for network geometry adaptation, and 2) the adapted network can accurately represent images with various contrasts (often acquired in high-dimensional imaging problems) by tuning only the latent space. Given a subject-specific high-resolution reference image $\mathbf{x}_p$ (e.g. from a reference T1w anatomical scan), the adaptation step can be formulated as:
\begin{equation}
\hat{\mathbf{w}}_p, \hat{\boldsymbol{\theta}}=\arg \min _{\mathbf{w}_p, \boldsymbol{\theta}}\left\|G_{\boldsymbol{\theta}}(\mathbf{w}_p)-\mathbf{x}_{p}\right\|_F^2+\alpha\left\|\boldsymbol{\theta}-\boldsymbol{\theta}_{p}\right\|_F^2,
\label{General_adaptation}
\end{equation}
\noindent where both the latents $\mathbf{w}_p$ and network parameters $\boldsymbol{\theta}$ are updated to minimize the representation error of the network for $\mathbf{x}_p$. The second term is introduced to penalize large deviation of the adapted network from the pretrained one. To reduce the optimization error and fully exploit the latent-space structures, we proposed to solve Eq.~(\ref{General_adaptation}) with a two-step algorithm, which alternates between: 
\begin{equation}
\hat{\mathbf{w}}_p=\arg \min _{\mathbf{w}_p}\left\|G_{\boldsymbol{\theta}_{p}}(\mathbf{w}_p)-\mathbf{x}_{p}\right\|_F^2,
\label{General_adaptation_1}
\end{equation}
\noindent and 
\begin{equation}
\hat{\boldsymbol{\theta}}=\arg \min _{ \boldsymbol{\theta}}\left\|G_{\boldsymbol{\theta}}(\hat{\mathbf{w}}_p)-\mathbf{x}_{p}\right\|_F^2+\alpha\left\|\boldsymbol{\theta}-\boldsymbol{\theta}_{p}\right\|_F^2.
\label{General_adaptation_2}
\end{equation}
The latents $\hat{\mathbf{w}}_p$ were solved by the ILO algorithm with fixed network parameters $\boldsymbol{\theta}_{p}$ (details provided below), and then the parameters were updated with fixed latents $\hat{\mathbf{w}}_p$.

This pretraining plus adaptation strategy produces an effective adapted GAN prior for which updating only the low-dimensional latents can accurately account for contrast variations, as demonstrated in Fig. \ref{Figure_adapt} using multicontrast brain MR images. The adaptive GAN can serve as a prior in many high-dimensional MR imaging applications, i.e., in this case introduced as a spatial constraint to enhance the subspace reconstruction.\\

\subsection{Algorithm}

With the adapted $\boldsymbol{G}_{\hat{\boldsymbol{\theta}}}(\cdot)$, we proposed to solve the problem in Eq.~\eqref{General_recon} using an alternating minimization algorithm. Specifically, the original reconstruction problem is decoupled into two subproblems, i.e., one StyleGAN inversion problem and the other GAN-constrained spatial coefficient update problem, which can be mathematically expressed as:\\

\noindent Subproblem (I): Update latent $\mathbf{w}$ by solving 
\begin{equation}
\{\hat{\mathbf{w}}_{t}^{i+1}\}=\arg \min _{\{\mathbf{w}_t\}} \sum_{t=1}^{N_t}\left\|(\hat{\U}^i \hat{\mathbf{V}})_t-G_{\hat{\boldsymbol{\theta}}}(\mathbf{w}_t)\right\|_F^2 ; 
\label{General_alter1}
\end{equation}
and \\

\begin{figure}[!t]
\centerline{\includegraphics[width=\columnwidth]{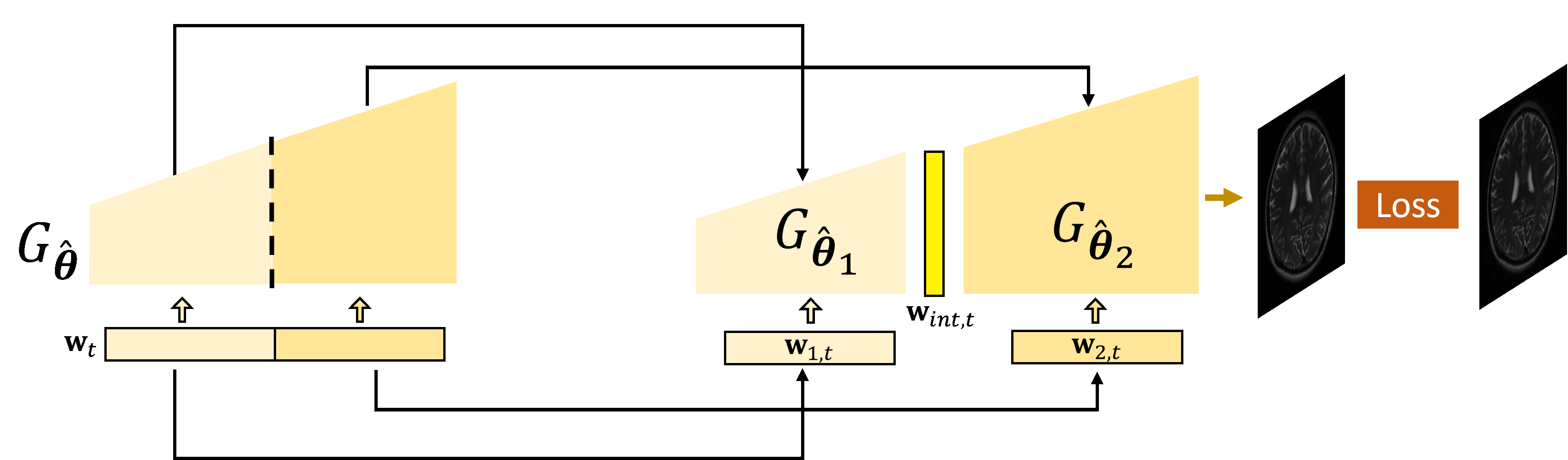}}
\caption{Illustration of the network partition in the ILO algorithm: The original StyleGAN $G_{\hat{\boldsymbol{\theta}}}$ can be divided into a cascade of two sub-networks, $G_{\hat{\boldsymbol{\theta}}_1}$ (lighter yellow) and $G_{\hat{\boldsymbol{\theta}}_2}$ (darker yellow), with corresponding latents $\w_{1,t}$ and $\w_{2,t}$. The output of $G_{\hat{\boldsymbol{\theta}}_1}$, $\w_{int,t}$ (originally hidden in $G_{\hat{\boldsymbol{\theta}}}$), can be viewed as intermediate latents and the input to $G_{\hat{\boldsymbol{\theta}}_2}$, which will be first updated in the GAN inversion process (see texts for details).}
\label{Figure_ILO}
\end{figure}

\noindent Subproblem(II): Update spatial coefficients $\mathbf{U}$ by solving
\begin{equation}
\begin{split}
\hat{\mathbf{U}}^{i+1}=& \arg \min _{\mathbf{U}} \left\|\mathbf{y}-\mathbf{A}( \mathbf{U} \hat{\mathbf{V}})\right\|_2^2 \\ & + \sum_{t=1}^{N_t} \lambda_{1,t} \left\| (\mathbf{U} \hat{\mathbf{V}})_t-G_{\hat{\boldsymbol{\theta}}}\left(\hat{\mathbf{w}}_{t}^{i+1}\right)\right\|_F^2+\lambda_2 R(\mathbf{U} \hat{\mathbf{V}}),
\label{General_alter2}
\end{split}
\end{equation}
where $i$ denotes the iteration number. Since the StyleGAN prior has already been adapted to the subject-specific reference image, for Subproblem (I) in Eq.~\eqref{General_alter1}, we minimized the image representation error with respect to the latents only. To address the optimization challenge for GAN inversion, we adapted the ILO algorithm considering the unique latent space structure in StyleGAN (similar to solving Eq.~\eqref{General_adaptation_1}). More specifically, instead of directly solving Eq.~\eqref{General_alter1} w.r.t. $\{\w_{t}\}$, the loss was first minimized w.r.t. the latents in an intermediate layer and then projected back to the range space of the previous layers. Our procedure is illustrated in Fig.~\ref{Figure_ILO}. 

Mathematically, we can denote the network as a composition of $G_{\hat{\boldsymbol{\theta}}}=G_{\hat{\boldsymbol{\theta}}_2} \circ G_{\hat{\boldsymbol{\theta}}_1}$, with $\hat{\boldsymbol{\theta}}_1$ and $\hat{\boldsymbol{\theta}}_2$ containing parameters for different partitions of the network (Fig.~\ref{Figure_ILO}). The multi-resolution latents in StyleGAN, $\w_t$, can then also be divided into two parts, $\w_{2,t}$ and $\w_{1,t}$ corresponding to the network partition above. With such a partition, we first minimize image representation error by updating $G_{\hat{\boldsymbol{\theta}}_2}(\cdot)$ w.r.t. a set of higher-resolution latents $\w_{2,t}$ and an intermediate layer output from the previous layers $\w_{int,t}$, i.e.,

\begin{equation}
\begin{split}
&\{\hat{\w}_{int,t}^{i+1}\},\{\hat{\mathbf{w}}_{2,t}^{i+1}\}
\\& =\argmin_{\{\w_{int,t}\},\{\w_{2,t}\}} \sum_{t=1}^{N_t}\left\|\hat{(\mathbf{U}}^i \hat{\mathbf{V}})_t-G_{\hat{\boldsymbol{\theta}}_2}(\w_{int,t},\w_{2,t})\right\|_F^2.\label{Inter_1}
\end{split}
\end{equation}

\begin{figure*}[!thb]
\centerline{\includegraphics[width=\textwidth]{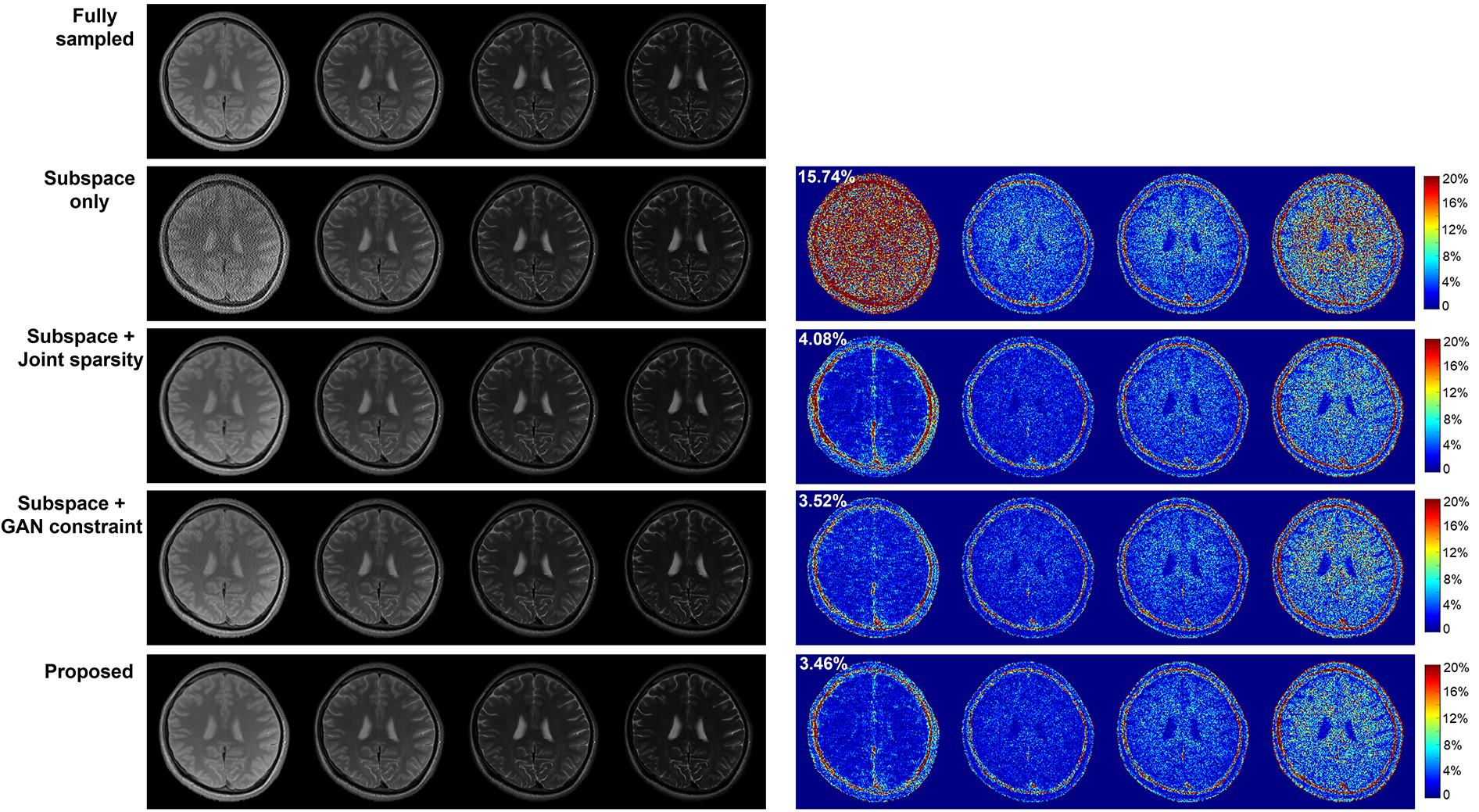}}
\caption{Reconstructed multi-TE images (Left panel) by different methods from an in vivo data set and the corresponding error maps (Right panel) at AF=4. Images for different TEs were shown in different columns while results for different methods in respective rows. The adaptive GAN representation worked as an effective spatial constraint that improved the subspace reconstruction (Row 4, Subspace+GAN constraint). The proposed method integrating all components outperformed the state-of-the-art subspace+sparsity method and achieved the lowest error (Last row), better visualized in the error images.}
\label{Results_GAN_T2}
\end{figure*}

\noindent  Subsequently, the update for the lower-resolution latents $\w_{1,t}$ was obtained by fitting $G_{\hat{\boldsymbol{\theta}}_1}$ to the intermediate $\{\hat{\w}_{int,t}^{i+1}\}$
\begin{equation}
\{\hat{\mathbf{w}}_{1,t}^{i+1}\}=\arg \min _{\{\mathbf{w}_{1,t}\}} \sum_{t=1}^{N_t}\left\| \hat{\mathbf{w}}_{int,t}^{i+1} -G_{\hat{\boldsymbol{\theta}}_1}(\mathbf{w}_{1,t})\right\|_F^2 . \label{Inter_2}
\end{equation}
\noindent This step essentially ``projects'' the intermediate layer output $\hat{\mathbf{w}}_{int, t}$ to the range space of $G_{\hat{\boldsymbol{\theta}}_1}$ by representing it with the lower resolution latents $\{\hat{\mathbf{w}}_{1,t}^{i+1}\}$. The resulting $\{\hat{\mathbf{w}}_{1,t}^{i+1}\},\{\hat{\mathbf{w}}_{2,t}^{i+1}\}$ can then be combined into an initialization for the problem in Eq.~\eqref{General_alter1} to further minimize the loss. In this work, to fully explore the multi-resolution latent structure in StyleGAN2, the network was divided into multiple partitions (with multiple intermediate latents) to break down the original challenging one step deep network inversion problem into multiple steps. Specifically, the minimization starts from a certain intermediate layer depending on the application and then backpropagated to the earlier layers (see the Experiments and Results section).

Additionally, we can assume that images sharing similar features, such as geometry and contrast, are likely to have similar latents (often demonstrated via ``style mixing'' experiments in the computer vision literature). Exploiting this assumption, we can mitigate potential overfitting to noise/artifacts by introducing a similarity constraint in the latent space across images with different contrasts. For example, we can introduce an $l_1$ ball constraint on the latents for adjacent ``time'' points, which can be mathematically formulated as:
\begin{equation}
\hat{\mathbf{w}}_{t}^{i+1}=\argmin _{\mathbf{w}_t
 \in \hat{\mathbf{w}}_{t-1}^{i+1} \oplus B_t(l_t)
} \left\|\hat{(\mathbf{U}}^i \hat{\mathbf{V}})_t-G_{\hat{\boldsymbol{\theta}}}(\mathbf{w}_t)\right\|_F^2 . \label{l1_ball}
\end{equation}
Here $\hat{\mathbf{w}}_{t-1}^{i+1} \oplus B_t(l_t)
$ denotes an $l_1$ ball with radius $l_t$ centered at $\hat{\mathbf{w}}_{t-1}^{i+1}$ from an adjacent ``time'' point (e.g., an image corresponding to a previous TE). The radius was determined empirically as hyperparameters (see the Discussion and Conclusion Section). Another strategy is to enforce latents at a certain resolution to be consistent when representing images across various contrasts (application dependent). 

For Subproblem (II), with the updated GAN representation, different regularization choices can be made for $R(\mathbf{U} \hat{\mathbf{V}})$ in Eq.~\eqref{General_alter2} in an application-specific context. This is essentially the well-studied regularized subspace reconstruction, but incorporating additional, reference-adapted network constraints. The initialization of alternating minization algorithm can be application specific, which will be discussed in the Experiments and Results section.

\subsection{Training and Other Implementation Details}
For the generative network, we used the StyleGAN2 architecture proposed in \cite{karras2020analyzing} with a latent space dimensionality of 512. Considering the complex-value nature of MRI data, we trained a magnitude image network and a phase image network separately. The magnitude network was trained on HCP database \cite{van2013wu} with 120,000 $\text{T}_1$-weighted $\&$ 120,000 $ 
\text{T}_2$-weighted images. The phase network was trained on NYU fastMRI database \cite{knoll2020fastmri} with 70,000 phase images after coil combination \cite{parker2014phase}.  The network
training, subject-specific adaptation and StyleGAN2 inversion were performed on a Linux server with NVIDIA A40 GPU and implemented in PyTorch 1.12.1. We used Adam optimizer \cite{kingma2014adam} with batch size 80 for training while other hyperparameters
remained unchanged as described in \cite{karras2020analyzing}. The subspace reconstruction and other methods for comparison were implemented in Matlab R2020b.

\section{Experiments and Results}
We evaluated the utility of the proposed method in two high-dimensional MR imaging applications: quantitative MR parameter mapping and MR spectroscopic imaging. Specific GAN adaptation considering the unique acquisition design in each imaging scenario, as well as modification to the overall formulation will be discussed. Improved reconstruction performance over state-of-the-art subspace reconstruction methods in each case will be demonstrated. All in vivo studies were performed with local IRB approval.
\subsection{Accelerated MR Parameter Mapping}
qMR provides quantitative tissue properties beyond traditional subjective and qualitative contrast-weighted images for diagnosis and disease characterization. One major challenge for clinical applications of qMR is the prolonged acquisition due to the need to acquire multiple images for parameter estimation. In this section, we demonstrate the effectiveness of our method for accelerating qMR (i.e., $\text{T}_2$ mapping) experiments.

In vivo $\text{T}_2$ mapping data were acquired using a multi-spin-echo sequence on a 3T scanner (Siemens Trio) using a 12-channel head coil. Specific acquisition parameters are: 16 echoes with $\mathrm{TE}_1=8.8 \mathrm{~ms}$ and echo spacing $\Delta \mathrm{TE}=8.8 \mathrm{~ms}$, $\mathrm{TR}= 4000 \mathrm{~ms}$, slice thickness $= 3$~mm, matrix size $= 192\times192$ and a FOV of 192$\times$192~mm$^2$. The acquired fully-sampled data were retrospectively undersampled using a 1D random phase encoding pattern at different acceleration factors (AFs). The central 12 $k$-space lines were acquired at all TEs for subspace determination \cite{peng2016accelerated,zhao2015accelerated}, and the central $k$-space was fully sampled at the first TE with coverage dependent on AFs, e.g., 48 for AF = 4, for coil sensitivity estimation by ESPIRiT \cite{uecker2014espirit}. The data and reconstruction workflow for this application is further illustrated in Fig.~S1 in the supplementary information.

To obtain a reference image for the proposed subject-specific GAN adaptation (Eq.~\eqref{General_adaptation}), we used a sum-of-squares (SoS) combination of images across all TEs from an initial subspace reconstruction. Based on our observation, the SoS image exhibits negligible aliasing even at high AFs. While additional high-resolution reference images can be acquired and used, e.g., a $\text{T}_1\text{w}$ image typically acquired in neuroimaging experiments, our strategy might alleviate the need of additional acquisitions. The coil sensitivity maps were integrated into the forward encoding model for reconstruction from multichannel data and the specific reconstruction formulation becomes: 
\begin{equation}
\begin{split}
&\hat{\mathbf{U}}, \{\hat{\mathbf{w}}_t\}= \arg \min _{\mathbf{U}, \{\mathbf{w}_t\}} \sum_{c=1}^{N_c} \left\|\mathbf{y}_c-\Omega( \mathbf{F} \mathbf{S}_c\mathbf{U} \hat{\mathbf{V}})\right\|_2^2 \\ & +\sum_{t=1}^{N_t} \lambda_{1,t} \left\|(\mathbf{U} \hat{\mathbf{V}})_t-\boldsymbol{\Phi}_t \odot\boldsymbol{G}_{\hat{\boldsymbol{\theta}}}(\mathbf{w}_t)\right\|_F^2 +\lambda_2 \|\mathbf{D} (\mathbf{U} \hat{\mathbf{V}})\|_{\mathbf{2}, \mathbf{1}},
\label{T2map_recon}
\end{split}
\end{equation}

\begin{figure}[!t]
\centerline{\includegraphics[width=\columnwidth]{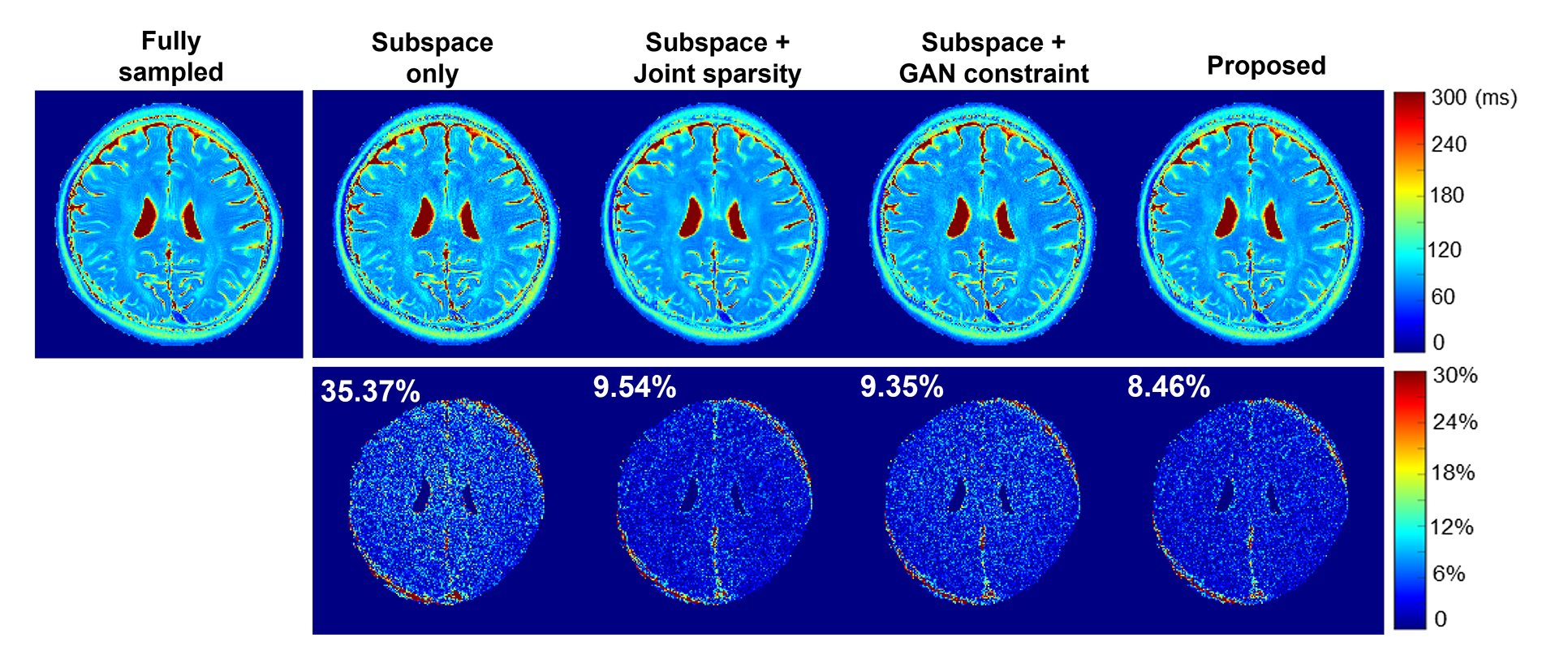}}
\caption{Estimated $\text{T}_2$ maps (Row 1) from the data in Fig.~3 and the corresponding error maps (Row 2) for different reconstruction methods ($\text{AF=4}$). The overall $\text{T}_2$ estimation errors are shown in the upper left corners of the images, which were calculated for the brain region only. As can be seen, reconstructions using joint sparsity constraint and GAN constraint yielded similar reconstruction errors. Integrating both sparsity and GAN constraints produced the best result. }
\label{Results_compare_T2}
\end{figure}

\begin{figure}[!t]
\centerline{\includegraphics[width=\columnwidth]{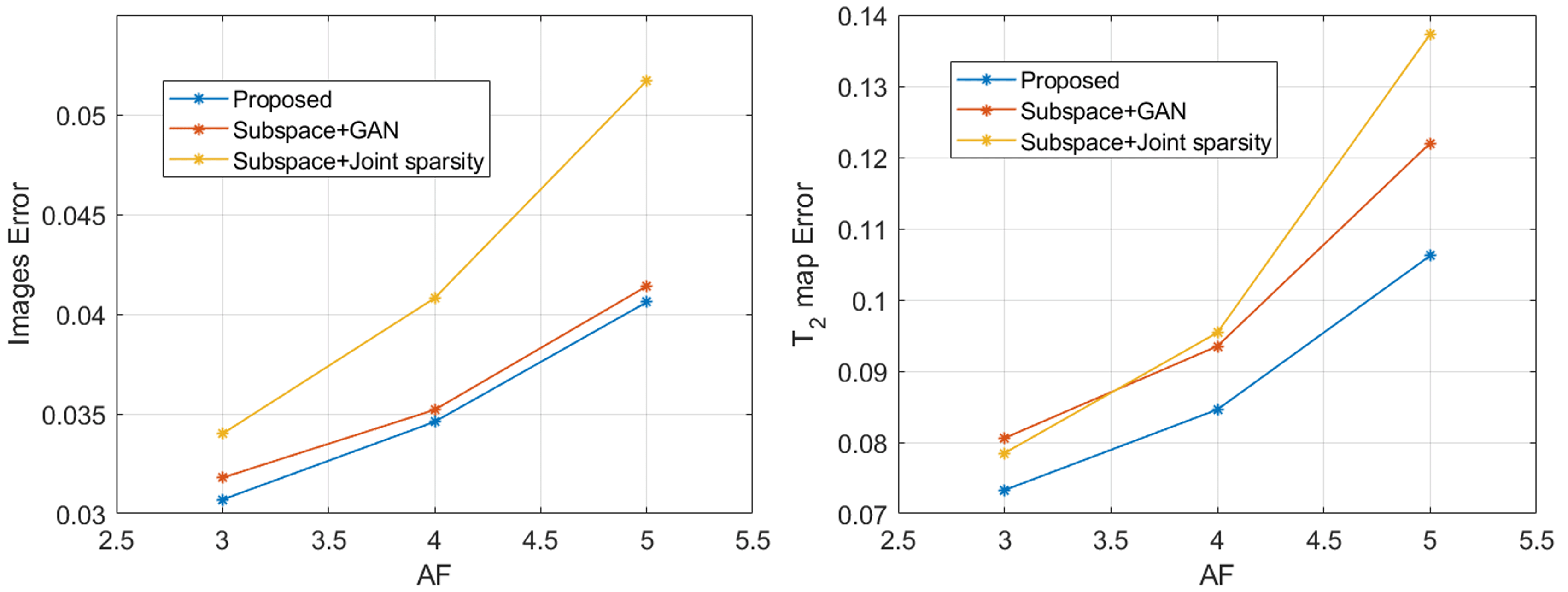}}
\caption{Reconstruction errors at different AFs for the $\text{T}_2$ mapping application: The proposed method shows lower errors across all AFs in both the reconstructed contrast-weighted images (left) as well as estimated $\text{T}_2$ maps (right).}
\label{Results_plot_T2}
\end{figure}

\noindent where $\Omega$, $\mathbf {F}$ and $\mathbf{S}_c$ denote the sampling operator, Fourier encoding matrix and sensitivity encoding matrix respectively. The additional regularization chosen here was a joint sparsity constraint with $\mathbf{D}$ a finite difference operator \cite{zhao2015accelerated}. 
We used magnitude network only and the phase $\{\boldsymbol{\Phi}_t\}$ came from a subspace reconstruction with joint sparsity constraint (Assuming the smooth phase from subspace reconstruction was good enough). The parameters $\{\lambda_{1,t}\}$ can be TE-dependent: we used $0.2$ for the last five TEs and $0.1$ for the other TEs in the reconstruction using subspace and adaptive GAN constraint alone. When using all three constraints, we simply chose $0.04$ for all $\lambda_{1,t}$ and achieved the best reconstruction results. For the joint sparsity contraint, we used a mild choice of $\lambda_2= 2e^{-7}$ ($\lambda_2= 1e^{-6}$ when using joint sparsity constraint alone). The algorithm converged after five iterations. More discussion on parameter selection can be found in the Discussion section.

When applying the proposed reconstruction, a similarity constraint in latent space was introduced for images at adjacent TEs in Subproblem (I) (as shown in Eq.~\eqref{l1_ball}). For initialization, we started the alternating minimization by first updating the latents (Eq.~\eqref{General_alter1}). More specifically,  we used a $\hat{\mathbf{U}}^0$ from a subspace reconstruction with a mild sparsity constraint, and then further updated the latents (before moving to Subproblem II) by solving the following problem each TE using $k$-space data directly to ensure data consistency: 
\begin{equation}
\hat{\mathbf{w}}_{t}^{0}=  \argmin _{\mathbf{w}_t
 \in \hat{\mathbf{w}}_{t-1}^{0}\oplus B_t(l_t)} \sum_{c=1}^{N_c} \left\|\mathbf{y}_{c,t}-{\Omega}_t [\mathbf{F} \mathbf{S}_c (\boldsymbol{\Phi}_t \odot G_{\hat{\boldsymbol{\theta}}}(\mathbf{w}_t))]\right\|_2^2.
\label{General_init1}
\end{equation}
\noindent Our StyleGAN2 network was trained with $192 \times 192$ images and $7$ layers in total to match the reconstruction resolution of the $\text{T}_2$ mapping data. The ILO algorithm started from the $5$th layer.

A set of multicontrast images reconstructed from data sampled with AF=4 are shown in Fig.~\ref{Results_GAN_T2}. Introducing either the sparsity constraint or the adapted GAN prior improved upon the subspace reconstruction, as expected, with the GAN prior producing a slightly lower image-domain error. The proposed method integrating both the constraints yielded the best reconstruction, as shown in the error images (right panel) and the overall relative $\ell_2$ errors. The $\text{T}_2$ estimation results corresponding to the same data (AF=4) are compared in Fig.~\ref{Results_compare_T2}, demonstrating effectiveness of the proposed method. More quantitative comparisons across different AFs are shown in Fig.~\ref{Results_plot_T2}. The proposed method consistently achieved lower errors at different AFs.
\subsection{High-Resolution MRSI}
\begin{figure}[!b]
\centerline{\includegraphics[width=0.90\columnwidth]{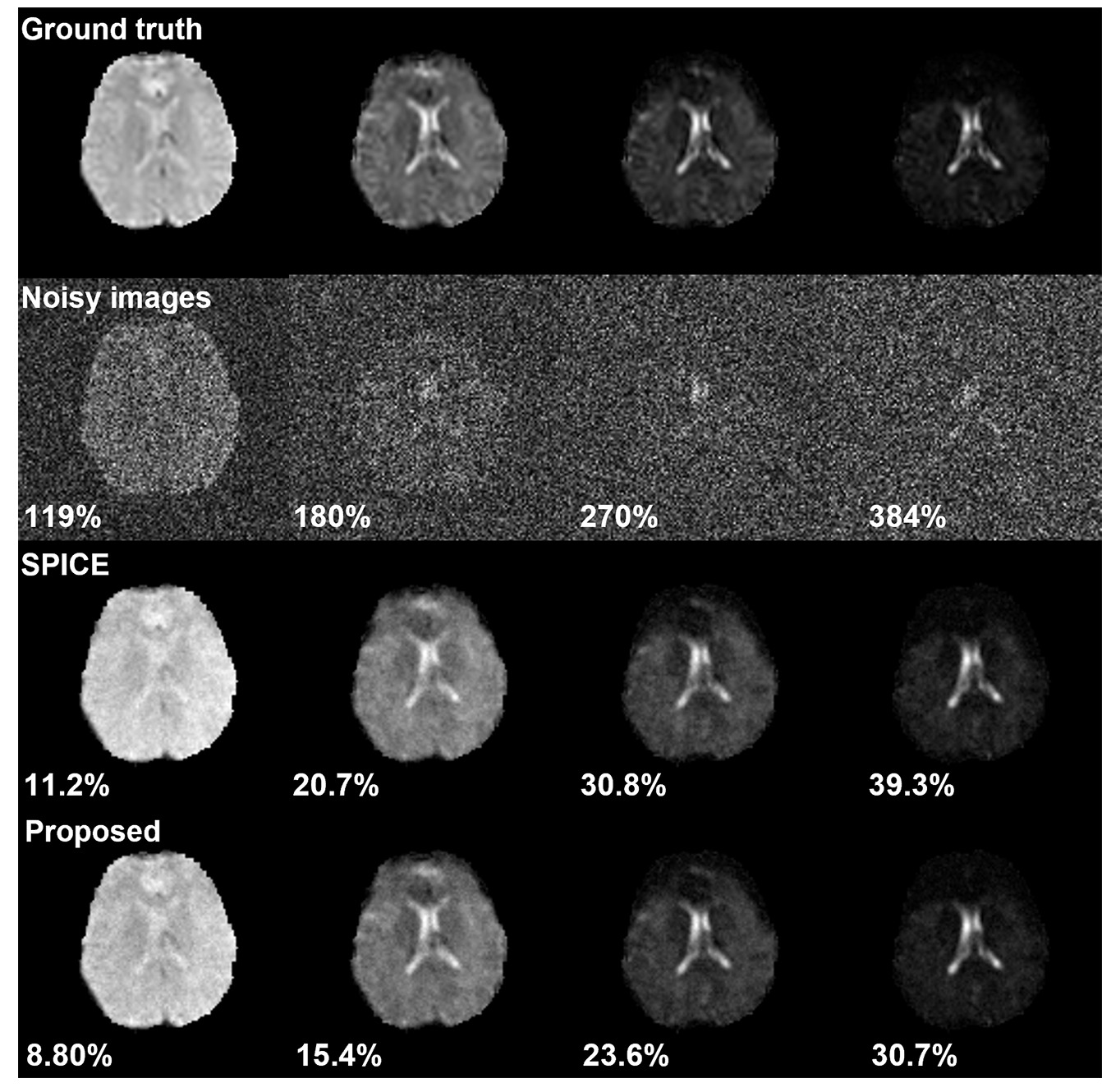}}
\caption{Simulation results for water MRSI reconstruction: (Row 1) Ground truth images simulated from interpolated water spectroscopic data ($128\times128\times12\times150$ spatiospectral encodings); 
(Row 2) Noisy images generated by adding Gaussian noise to ground truth; (Rows 3 and 4) Reconstruction results from the SPICE subspace reconstruction and proposed method (Proposed). The proposed method produced better contrast, a higher SNR and a lower reconstruction error.
}
\label{MRSI_simu}
\end{figure}
In this section, we demonstrated the effectiveness of the proposed method on another high-dimensional imaging problem, i.e., MRSI. The goal here is to evaluate our method's utility for SNR-enhancing reconstruction from noisy, high-resolution MRSI acquisitions, benchmarking against the state-of-the-art subspace reconstruction (SPICE).
\begin{figure*}[!tbh]
\centerline{\includegraphics[width=0.90\textwidth]{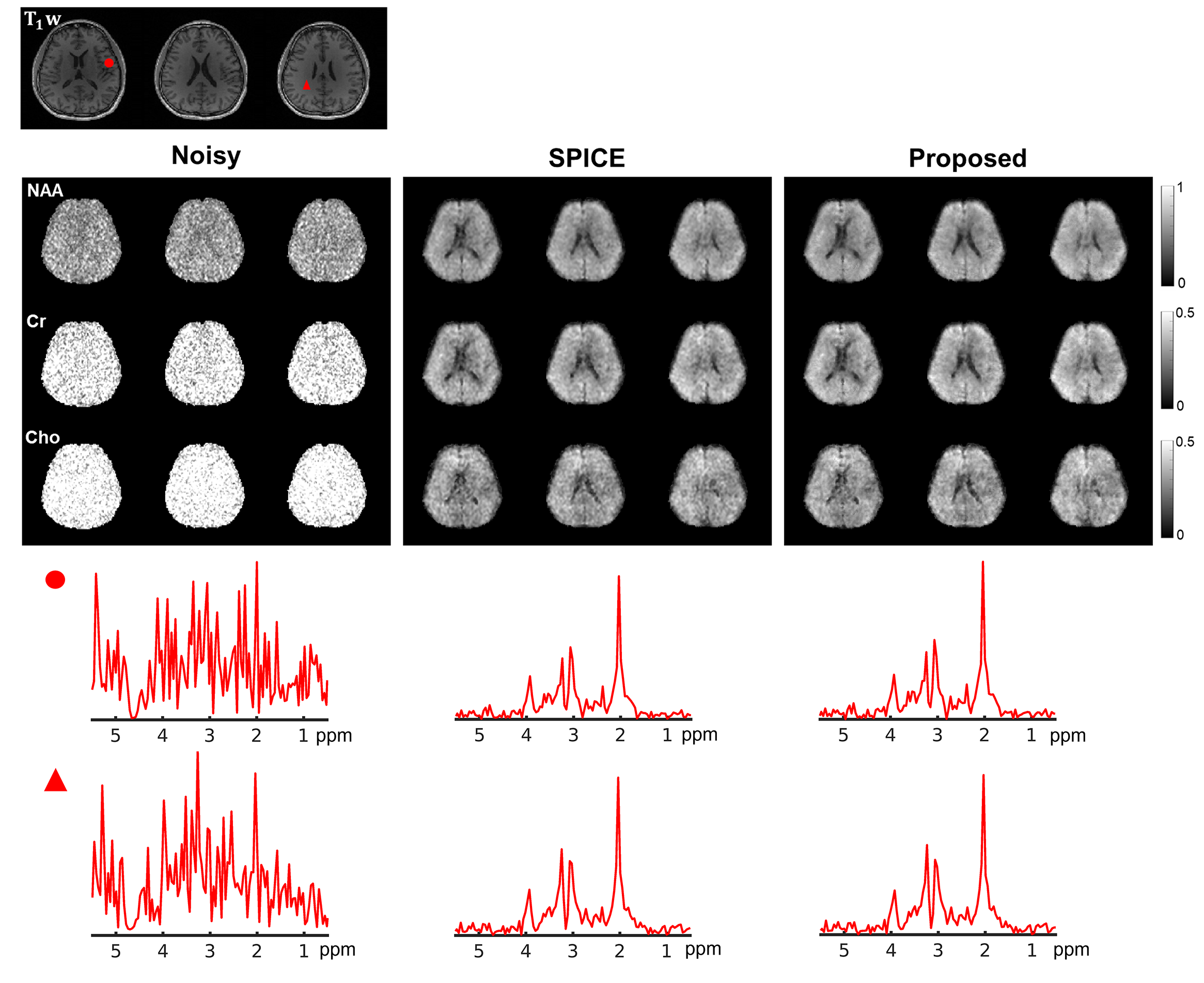}}
\caption{In vivo MRSI reconstruction: (Row 1) $\text{T}_1$-weighted anatomical images ($\text{T}_1$w, serving as the reference for network adaptation); (Rows 2-4) Metabolite maps for the corresponding slices from noisy data (left three columns), SPICE (middle three columns) and the proposed method (right three columns), respectively. Both SPICE and the proposed method produced significant SNR enhancement for this high-resolution acquisition, revealing tissue specific metabolite variations, but better contrast and less artifacts can be observed in the proposed reconstruction. The last two rows show two voxel spectra from respective methods (voxel locations marked in the $\text{T}_1$w images). Similar spectral quality was achieved by the SPICE and proposed methods (not surprising as both methods used the same subspace).}
\label{MRSI_results1}
\end{figure*}

As it is rather difficult to obtain ``reference'' high-SNR data for high-resolution MRSI, we performed both numerical simulations (with ground truth) and in vivo experiments for evaluation. In vivo MRSI data were acquired from healthy volunteers using a fast sequence described in Ref.~\cite{wang2022high}, with $\text{TR/TE}=1100/30 \mathrm{~ms}$, $\text{matrix size}=64 \times 64 \times 12$,
$\text{FOV}=220 \times 220 \times 64\mathrm{~mm}^3$, spectral bandwidth (BW) of 1.67kHz and 320 echoes (spectral encodings). Our sequence also generated a set of interleaved high-resolution, high-SNR water spectroscopic data, which is an excellent data for simulation-based evaluation, particularly for the adaptive GAN representation for different contrasts. For simulations, we first interpolated the water spectroscopic images to a grid of $128 \times 128 \times12 \times 150$ (where 150 is the number of samples along the FID dimension) and used them as the ground truth to simulate noisy water MRSI data (Fig.~6). The noisy metabolic MRSI data from the same sequence were used to evaluate the proposed method for metabolite spatiospectral reconstruction. In MRSI experiments, a high-resolution, anatomical scan is typically acquired and can be used as a reference for GAN adaptation. Here, a 3D $\text{T}_1\text{w}$ image (from an MPRAGE scan) was acquired.

The StyleGAN2 network was pretrained using the same HCP data but resized to $128\times 128$ (as MRSI data have a lower resolution than $\text{T}_2$ mapping) and skull stripped. The pretrained network was first adapted to the $\text{T}_1\text{w}$ reference image to generate $\boldsymbol{G}_{\hat{\boldsymbol{\theta}}}(\cdot)$ for the proposed reconstruction. The phase network $\boldsymbol{G}_{\hat{\boldsymbol{\theta}}_\phi}(\cdot)$ pretrained on NYU fastMRI data was introduced to account for the complex-valued data. Considering that a) the MRSI data have a significantly larger number of ``time points'' than parameter mapping and b) the images at individual time points (especially later echoes) are much noiser, instead of representing images at different echoes using the adapted StyleGAN2, we proposed to use it to represent the spatial coefficient maps (for subspace reconstruction) as they can be treated as "images" with different contrasts. The specific formulation can be written as:
\begin{equation}
\begin{split}
& \hat{\mathbf{U}}, \{\hat{\mathbf{w}}_r\},\{\hat{\mathbf{w}}_{\phi,r}\}= \arg \min _{\mathbf{U}, \{\mathbf{w}_r\},\{\mathbf{w}_{\phi,r}\}} \left\|\mathbf{y}-\Omega [\mathbf {F} \mathbf{B} \odot (\mathbf{U} \hat{\mathbf{V}})]\right\|_2^2 \\ & + \sum_{r=1}^{R} \lambda_{1,r} \left\|{\U}_r-G_{\hat{\boldsymbol{\theta}}}(\mathbf{w}_r) \odot G_{\hat{\boldsymbol{\theta}}_\phi}(\mathbf{w}_{\phi,r})\right\|_F^2 +\lambda_2 \|\mathbf{D_w} (\mathbf{U} \hat{\mathbf{V}})\|_F^2,
\label{MRSI_recon}
\end{split}
\end{equation}

\noindent where ${\U}_r$ is the $r_{th}$ column in $\U$ (spatial map of each coefficient) and $R$ is the number of columns (aka the rank), $\mathbf{B}$ models the field inhomogeneity phase effects and $\mathbf{D_w}$ denotes an edge-weighted finite difference operator. This led to a slightly different Subproblem I than the previous application example, where the latents for magnitude network ($\{{\mathbf{w}}_r\}$) and phase network ($\{{\mathbf{w}}_{\phi,r}\}$) were determined alternatively. We performed a SPICE-based subspace reconstruction with a mild edge-preserving regularization as the initialization for $\hat{\mathbf{U}}^0$. The high-resolution latents for the StyleGAN2 (inputs for the last layer) were kept unchanged (after adaptation in Eq.~\eqref{General_adaptation_2}) during the GAN inversion subproblem to alleviate overfiting to noise considering the low SNR. The effects of changing latents at different resolution are further illustrated in the Discussion. We chose $\lambda_{1,r}=1.6$ and $\lambda_{2}=0.3$ for proposed method ($\lambda_{2}=0.6$ for the SPICE reconstruction with a similar data consistency level). The ILO started from the $4$th layer of the network.

Reconstruction from the simulated noisy water images supported the idea of using adaptive GAN to represent the spatial coefficient maps. The subspace for water reconstruction was obtained by applying an SVD to the ``ground truth'' water spectroscopic images. As shown in Fig.~\ref{MRSI_simu}, the proposed method achieved a significantly lower reconstruction error and noticeably better ``contrast recovery'' than the SPICE reconstruction. Impressive SNR enhancement was achieved by both. Fig.~\ref{MRSI_results1} shows a set of metabolite spatiospectral reconstruction from an in vivo MRSI acquisition. The metabolite maps from the proposed reconstruction exhibit a higher SNR and clearer contrast compared to the noisy data and subspace reconstruction (SPICE). Note that increasing the spatial regularization parameter for the SPICE reconstruction can achieve a similar level of SNR enhancement but at the expense of oversmoothing. The regularization parameters were tuned to a similar data consistency level, matching noise energy for both methods. Therefore, the proposed method provided a better trade-off in SNR and resolution by incorporating an adaptive GAN-based spatial constraint. The subspace used for the MRSI reconstruction was pre-estimated and adapted to this specific data (using the higher-SNR portion) as described in  \cite{lam2020ultrafast,guo2021simultaneous,peng2018simultaneous}.

\section{Discussion}

\begin{figure}[!t]
\centerline{\includegraphics[width=0.9\columnwidth]{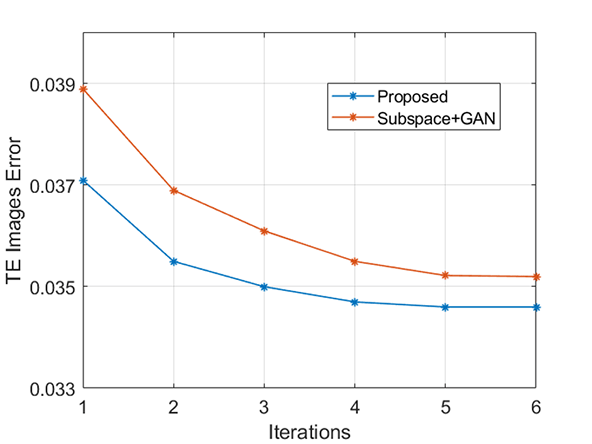}}
\caption{ The effects of iteration for our proposed algorithm: Reconstruction errors of the multi-TE images for a $\text{T}_2$ mapping data set w.r.t. the number of iteration (AF=4). The error changes substantially for the first 2-3 iterations, but minimally after 5, with and without integrating the adaptive GAN and joint sparsity constraints.
 }
\label{Convergency}
\end{figure}
We presented a new adaptive generative model based spatial constraint for subspace reconstruction of high-dimensional imaging data. One unique advantage of our proposed method is that the subject-specific adaptation step allows our GAN model to be flexibly adapted to different imaging contexts, even without high-quality, application-specific training data. We validated that the subject-specific, adapted GAN (to an application-specific reference image) can be accurate representation of images from the same subject with different contrasts (a common feature in high-dimensional imaging) by updating latent variables alone. This circumvented the overfitting issue in methods updating both latents and network parameters during reconstruction, such as IAGAN. We demonstrated the effectiveness of our method in two application examples.

There are several points worth discussing. The first is the selection of regularization parameter(s) (and other hyperparameter(s)), particularly when our adaptive GAN prior is combined with other regularization functional. To alleviate possible overregularization/oversmoothing introduced by other handcrafted constraints, we chose a small $\lambda_2$ in this work for a mild additional regularization effect, and our proposed GAN constraint played a more important role. For $\{\lambda_{1,t}\}$, selection was based on a combination of discrepancy principle and visual inspection. We also observed that in our $\text{T}_2$ mapping experiments, similar $\{\lambda_{1,t}\}$ worked well across different AFs, and a relatively large range of $\{\lambda_{1,t}\}$ can result in similar reconstruction results. One advantage is that $\{\lambda_{1,t}\}$ can be ``time'' specific. More specifically, when using adaptive GAN only to constrain the subspace reconstruction, a 2$\times$ larger regularization parameter for the last five TEs (due to relatively lower SNRs) can slightly improve the $\text{T}_2$ mapping reconstruction. Different combinations of $\{\lambda_{1,t}\}$ may be further explored. Similarly, the determination of the radius $l_t$ that imposes the $l_1$ ball constraint (Eq.~\eqref{l1_ball})  can be ``time'' specific and further explored. The second is the initialization step. We observed that the reconstruction can be substantially affected by initialization which should be carefully chosen for different applications. For our $\text{T}_2$ mapping experiments, starting the iterations using $\hat{\mathbf{U}}^0$ from a subspace reconstruction (with a mild joint sparsity constraint) may introduce potential bias. Therefore, we proposed to further update the latents using the data consistency loss described in Eq.~\eqref{General_init1} which led to a better performance. For MRSI, we found that initializing the algorithm with $\hat{\mathbf{U}}^0$ from a initial subspace reconstruction worked well. 

Another issue is the reconstruction speed. Currently, we used five outer iterations (Fig.~\ref{Convergency}) that  achieved empirical convergence and satisfactory reconstruction results. Each iteration needs about 40 minutes (can be application dependent). The most computationally expensive step is
Subproblem (I), where multiple gradient descent updates are required to explore the intermediate latent space. Currently, we solved the latents TE by TE for MR parameter mapping, for which more efficient implementations should be considered. Other methods that better exploit parallel computation to simultaneously update the latents for all TEs and even circumvent network inversion will be investigated.

\begin{figure}[!t]
\centerline{\includegraphics[width=\columnwidth]{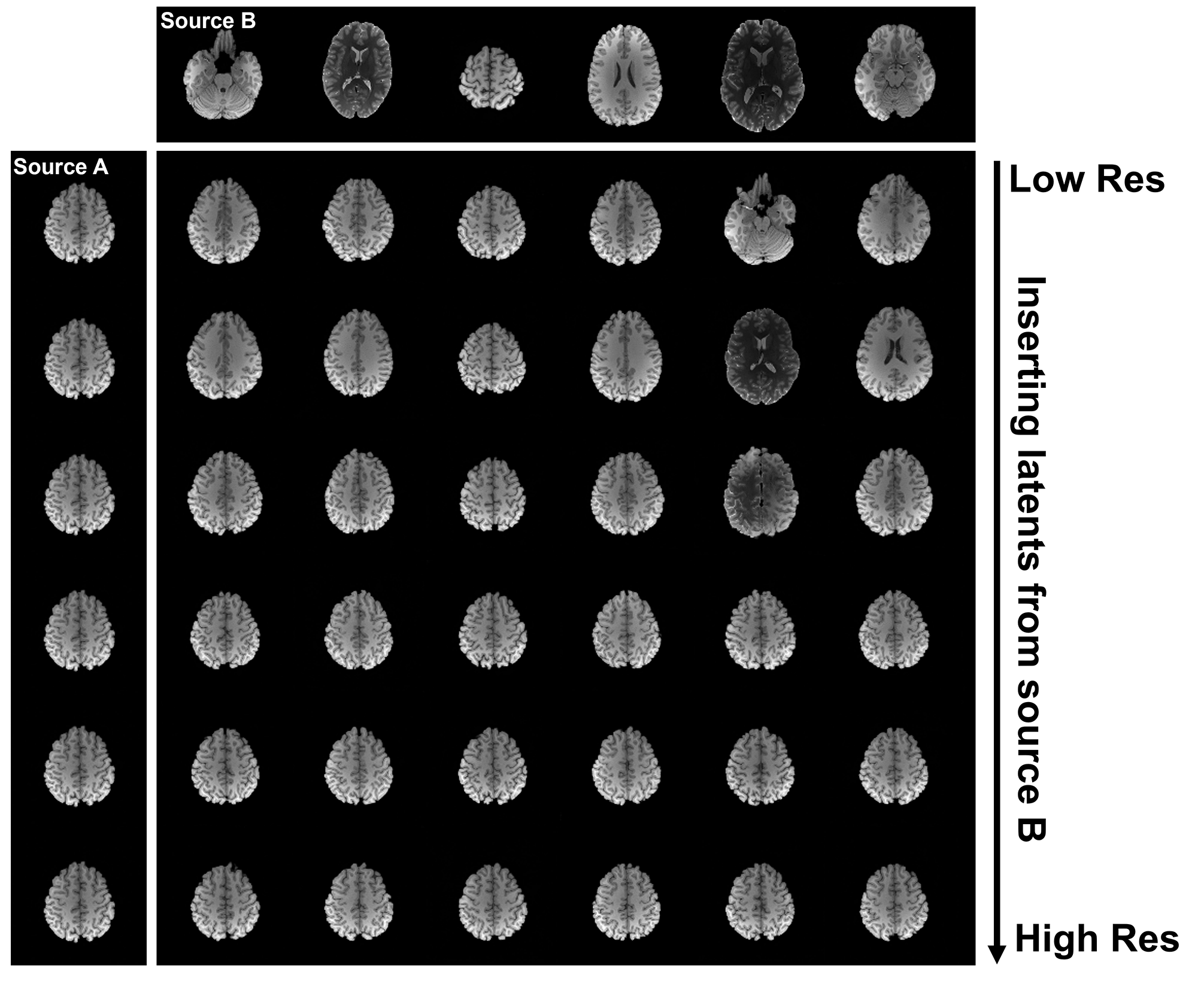}}
\caption{A style mixing experiment for our adapted StyleGAN2: A single image (Source A) and a set of images (Source B) were generated from the network; The rest of the images were generated by replacing a specific subset of latents in Source A by corresponding latents from Source B. Specifically, for each row, images in different columns were obtained by taking the same level of latents from different images in Source B. From the top to bottom rows, different levels of latents from Source B were used (for the same image A), from low-resolution to high-resolution subsets, respectively. As can be seen, lower-resolution latents from Source B are more related to global features such as shape and contrast, while higher-resolution latents lead to more subtle changes in fine details and less/negligible contrast and geometry variations. 
 }
\label{Stylemixing}
\end{figure}

While our work and previous work \cite{kelkar2021prior} using GAN for image reconstruction have demonstrated that imposing constraints on latents can help better utilize prior information from reference images and reduce overfitting to artifacts, an important issue that presents unique research opportunities is the properties of latents (feature) space. Here, we further investigated the latent space structure through a style-mixing experiment on the adapted StyleGAN2 (after pre-training and network parameter adaptation to experimentally acquired $\text{T}_1\text{w}$ image). Fig.~\ref{Stylemixing} demonstrated that manipulating low-resolution latents can introduce more global and significant changes in structure and contrast of the generated images than changing the high-resolution latents. This is consistent to us keeping the high-resolution latents unchanged for our MRSI reconstruction (lower-resolution data and avoiding overfitting to noise). In the meantime, it can be observed that for the current StyleGAN representation, different types of image features (e.g. contrast and geometry) are still somewhat entangled, and it is challenging to control certain semantic features by changing latents at a specific scale. Future work can focus on developing disentangled representation to enable more explicit control of different image features, which will be desirable for high-dimensional MR imaging problems, since in many situations, only parts of image features are changing across specific dimensions.  

Recent development of generative models may also be helpful to improve the accuracy and efficiency of image representation, and thus improve the reconstruction performance. Transformer-based GAN has been proposed \cite{jiang2021transgan} to better utilize the 
global correspondence of image features through self-attention mechanism. Korkmaz et al. \cite{korkmaz2021deep} demonstrated that generative vision transformers were more robust to artifacts and achieved better performance than CNN-based generative models when worked as a deep image prior (without pretraining) \cite{ulyanov2018deep}. Diffusion model is an emerging approach to generate
high-quality images. Diffusion model training is more stable than GAN and can obtain better sample quality over the state-of-the-art GANs \cite{dhariwal2021diffusion,
ho2022cascaded}. While diffusion models have demonstrated its potential for image reconstruction \cite{dar2022adaptive,song2022solving}, many opportunities remain on how to incorporate diffusion models into high-dimensional MR imaging problems \cite{levac2023conditional}.

\section{Conclusion}
We proposed a novel reconstruction method for high-dimensional MR imaging that integrated subspace modeling and an adaptive generative-network-based image prior. The proposed adaptive generative model served as an accurate representation of images from the same subject with different contrasts and an effective spatial constraint for subspace reconstruction. The complementary powers of subspace, generative image constraint, and sparsity regularization produced improved performance over state-of-the-art subspace methods in two application examples. We believe our work offers a new perspective on high-dimensional image reconstruction via integrating learning-based spatial priors and low-dimensional modeling.

\bibliographystyle{IEEEtrans}
\bibliography{ref}

\end{document}


\section*{Supplementary Figures}
\noindent
\setcounter{figure}{0}
\renewcommand{\thefigure}{S\arabic{figure}}

\begin{figure}[thb!]
\centerline{\includegraphics[width=0.81\columnwidth]{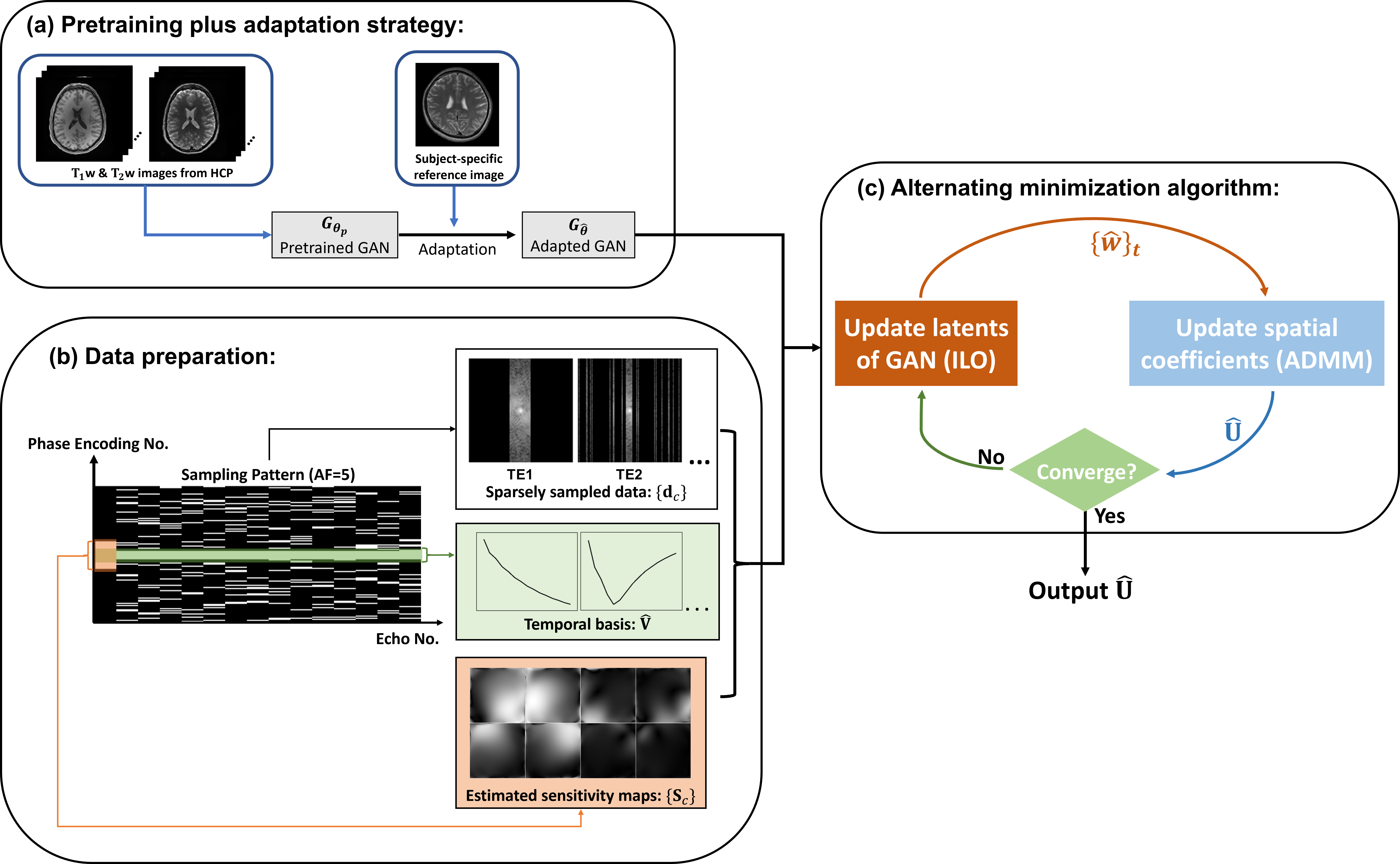}}
\caption{An illustration of the reconstruction workflow for accelerated $\text{T}_2$ mapping: (a) Subject-specific network representation $G_{\hat{\boldsymbol{\theta}}}$ was obtained by adapting a StyleGAN2 pretrained on HCP data and used as image-domain constraints for different contrast-weighted images; (b) Sparsely sampled data were acquired; temporal basis and sensitivity maps (first TE) were estimated from partial data; (c) An alternating minimization algorithm that updates the latents and spatial coefficients alternatively was developed to solve the optimization problem using the multichannel, sparse $k$-space data.}
\label{T2map_sup}
\end{figure}

\begin{figure}[thb!]
\centerline{\includegraphics[width=0.83\columnwidth]{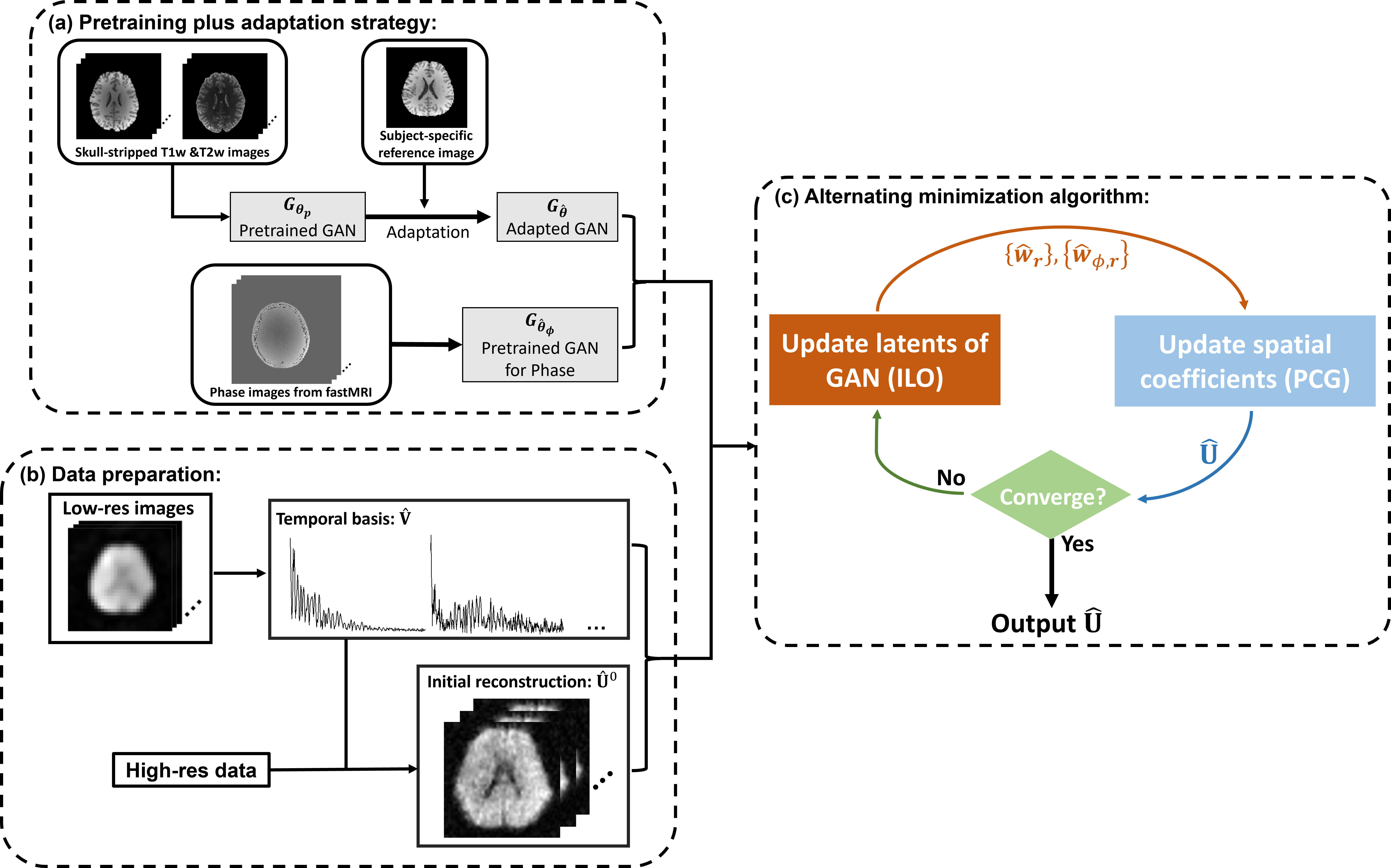}}
\caption{An illustration of the reconstruction workflow for the high-resolution MRSI problem: (a) The same pre-training plus adaptation strategy was used, producing a constraint for the spatial coefficients. An additional phase network pretrained on NYU fastMRI data was introduced for phase constraint; (b) Temporal basis was pre-estimated from subject-specific low-resolution data (higher SNR). An initial subspace reconstruction was performed for algorithm initialization; (c) A similar alternating minimization algorithm was used to solve the reconstruction problem. Latents for magnitude and phase networks were updated alternatively during the styleGAN2 latent update step.}
\label{MRSI_sup}
\end{figure}